\documentclass[useAMS,usenatbib]{mn2e}
\usepackage{graphicx}

\def\ltorder{\mathrel{\raise.3ex\hbox{$<$}\mkern-14mu\lower0.6ex\hbox
{$\sim$}}}
\def\gtorder{\mathrel{\raise.3ex\hbox{$>$}\mkern-14mu\lower0.6ex\hbox
{$\sim$}}}

\title[Galaxies as Fluctuations in the Ionizing Background Radiation]
  {Galaxies as Fluctuations in the Ionizing Background Radiation at Low
  Redshift}
\author[Suzanne M. Linder et al.]
  {Suzanne M. Linder$^1$, Roland Gunesch$^2$, Jonathan I. Davies$^1$,
Maarten Baes$^{1,3}$,\newauthor
Rhodri Evans$^1$, Sarah Roberts$^1$, Sabina Sabatini$^1$,
and Rodney Smith$^1$ \\
$^1$Cardiff University, Department of Physics and Astronomy, Queen's
Buildings, P.O. Box 913, Cardiff CF24 3YB, Wales, UK\\ $^2$Mathematisches
Institut, Universit\"at Leipzig, Augustusplatz 10-11, D-01409 Leipzig,
Germany\\ $^3$Sterrenkundig Observatorium, Universiteit Gent, Krijgslaan 281-S9,
B-9000 Gent, Belgium
      }
\date{Released 2002 Xxxxx XX}

\pagerange{\pageref{firstpage}--\pageref{lastpage}} \pubyear{2003}

\def\LaTeX{L\kern-.36em\raise.3ex\hbox{a}\kern-.15em
    T\kern-.1667em\lower.7ex\hbox{E}\kern-.125emX}

\begin{document}

\label{firstpage}

\maketitle

\begin{abstract}
Some Lyman continuum photons are likely to escape from most
galaxies, and these can play an important role in ionizing gas around and
between galaxies, including gas that gives rise to Lyman alpha absorption.
Thus the gas surrounding
galaxies and in the intergalactic medium will be exposed to varying amounts of
ionizing radiation depending upon the distances, orientations, and luminosities
of any nearby galaxies.
The ionizing background can be recalculated at any point within
a simulation by adding the flux from the galaxies
to a uniform quasar contribution.
Normal galaxies are found to almost always make some
contribution to the ionizing background radiation at redshift zero, as
seen by absorbers and at
random points in space.  Assuming that $\sim 2$ percent of ionizing photons escape
from a galaxy like the Milky Way, we find that normal
galaxies make a contribution of
at least 30 to 40 percent of the assumed quasar background.
Lyman alpha absorbers with a wide range of
neutral column densities are found to be exposed to a wide range of ionization
rates, although the distribution of photoionization rates for absorbers is found to
be strongly peaked.  On average, less highly ionized absorbers are found to arise farther
from luminous galaxies, while local fluctuations in the ionization rate
are seen around galaxies having a wide range of properties.
\end{abstract}


\section{Introduction}

The extragalactic background of Lyman continuum photons plays an
important role in ionizing the many absorption line systems seen
shortward  of Ly$\alpha$ emission in
quasar spectra, including those seen at low redshifts using the
ultraviolet capabilities of the Hubble Space Telescope (Bahcall
et al.~1996).
The ionizing background intensity determines the neutral gas
fraction in Ly$\alpha$ forest absorbers and the ion ratios seen in
metal absorption line systems.	Furthermore, understanding the ionizing
background is important for developing theories for the formation and
evolution of galaxies and the Ly$\alpha$ forest and possibly for finding
the baryonic mass contained within galaxies and the intergalactic medium.
Interesting questions remain as to how much of this background is
contributed by galaxies and what role galaxies play in ionizing gas around
and between them that is detected as Ly$\alpha$ absorption.

The
only largely neutral Ly$\alpha$
absorbers are the damped systems which have neutral
hydrogen column densities $N_{HI} > 10^{20.3}$ cm$^{-2}$.
Ionized absorbers include some Lyman limit systems ($N_{HI} > 10^{17.2}$ cm$
^{-2}$) and Ly$\alpha$ forest absorbers which have lower neutral column
densities.  Lyman limit systems are thought to arise around galaxies
(Bergeron \& Boiss\'e 1991; Steidel 1995),
while Ly$\alpha$ forest absorbers have been
detected which are as weak as $N_{HI}
\sim 10^{12}$ cm$^{-2}$.   Although many of these may be associated with the
smallest amounts of intergalactic gas, including that in void
regions (Dav\'e et al.~1999; Stocke et al.~1995;
Shull et al.~1996; Penton, Stocke, \& Shull
2002),
at least some
stronger forest absorbers are found near luminous galaxies.
In particular
Ly$\alpha$ absorption is almost always detected at a similar redshift
to a galaxy which is found within $\sim 200$ kpc of a quasar line of
sight
(Bowen, Blades, \& Pettini 1996;
Lanzetta et al.~1995a;
Le Brun, Bergeron, \& Boiss\'e 1996;
Chen et al.~1998; 2001).
Bowen, Pettini, \& Blades (2002) have shown recently that while nearby Ly$\alpha$
absorbers are difficult to match with particular observed galaxies, $N_{HI}$
is correlated with the local density of detected luminous galaxies.

In carefully observed spiral galaxy discs the neutral column density
falls off slowly with radius over most of the extents as seen with 21 cm
HI observations, but then a rapid truncation is seen.  Bochkarev \&
Sunyaev (1977) first suggested that a truncation would occur in a spiral
disc at a sufficient radius where the HI column becomes ionized by an
extragalactic background.  Disk edges have been modelled also by Maloney
(1993), Dove \& Shull (1994a), and Corbelli \& Salpeter (1993).	 More
recently ionized gas has been detected in H$\alpha$ emission using a
Fabry-Perot `staring technique' (Bland-Hawthorn et al.~1994) beyond the
HI edges of several nearby galaxies
(Bland-Hawthorn, Freeman \& Quinn 1997; Bland-Hawthorn 1998).

The intensity of the ionizing background radiation has been measured
using the `proximity effect', or the paucity of absorption lines close
to a quasar emission redshift, at low redshift first by Kulkarni \& Fall
(1993) and more recently by Scott et al.~(2002) who find $J(912$\AA$)=
7.6^{+9.4}_{-3.0}\times 10^{-23}$ erg cm$^{-2}$ s$^{-1}$ Hz$^{-1}$
sr$^{-1}$, or a frequency and direction averaged ionization rate of
$1.9 \times 10^{-13}$ s$^{-1}$ at redshift
$z<1$.  Some evidence for redshift evolution in the background
is seen.  Other methods for estimating this intensity give
generally consistent
results typically within the lower end of their uncertainty range,
including limits on H$\alpha$ emission from high-latitude
galactic clouds (Vogel 1995; Tufte, Reynolds, \& Haffner 1998; Vogel et al.~2002) and
extragalactic HI clouds (Stocke et al.~1991; Donahue, Aldering \& Stocke
1995) and estimates from the HI galaxy disc edges (Maloney 1993; Dove
\& Shull 1994a; Corbelli \& Salpeter 1993).

Using quasar spectra and considering reprocessing of photons by the
intergalactic medium, Haardt \& Madau (1996), Dav\'e et al.~(1999),
and Shull et al.~(1999) have calculated the history of the
intensity of the ionizing
background down to low redshifts.  Their values are approximately
consistent with the measurements above,	 so that quasars make an important
and possibly dominant contribution to the
ionizing background even at low redshifts.  While the number density of
quasars is low at low redshifts, the universe becomes optically thin
to ultraviolet photons at redshifts $z \ltorder 2$ (Haardt \& Madau 1996)
such that ultraviolet photons emitted at $z\sim 2$ are likely to survive
without reprocessing until $z\sim 0$.  In contrast, the ionizing spectrum
at higher redshifts is modified by absorption and reemission (Fardal,
Giroux, \& Shull 1998).

What contribution might galaxies make to the ionizing background at
low redshifts?	Some suggestions have been made that insufficient numbers
of ionizing photons (less than one percent) escape from galaxies for an
important contribution to the ionizing background
(Deharveng et al.~1997;
Henry 2002), though others measure (Bland-Hawthorn \& Maloney 1999; Leitherer
et al.~1995; Goldader et al.~2002; Hurwitz, Jelinsky, \& Van Dyke Dixon 1997) and
model (Dove \& Shull 1994b; Dove, Shull \& Ferrara 2000) higher escape
fractions between three and ten percent.  Giallongo, Fontana, \& Madau (1997),
Shull et al.~(1999) and
Bianchi, Cristiani, \& Kim (2001) find that star-forming galaxies could even
dominate
the ionizing background if at least a few percent of the ultraviolet
photons escape.  Some ionizing photons are likely to escape from most galaxies.
Bland-Hawthorn (1998) suggests that
the gas detected beyond the HI disc edge in several spiral galaxies
is ionized by stellar populations within the galaxies, as the emission
measures are stronger than those predicted (Maloney 1993; Dove \& Shull
1994a) for an extragalactic background.

Given that many stronger Ly$\alpha$ absorbers are found close to galaxies,
it is possible that stellar populations within galaxies make some
contribution to the ultraviolet photons that ionize any nearby absorbers.
Thus what is measured as ionizing background radiation may vary in
intensity with location, depending upon the galaxy clustering environment
or the properties, such as brightness and extinction behaviour, of any
nearby galaxies.
The method for simulating a fluctuating ionizing background is
described in Section 2, while the
resulting fluctuations are discussed in Section 3.  Results of varying 
model parameters are discussed in Section 4, and
Section 5 describes the relationship of the ionization rate fluctuations
to the properties and locations of galaxies.  The value
of $H_0$ is assumed to be 100 km s$^{-1}$ Mpc$^{-1}$.

\section{Method and Simulations}

The simulation used here is an updated version of that first described in
Linder (1998), where in each case here 12590 clustered galaxies are placed
in a cube with an edge of 28.9 Mpc (except where this edge is adjusted at
redshift one).  Ly$\alpha$ absorbers
arise in gas within and extending from galaxy discs, which are modelled as in
Charlton, Salpeter, \& Hogan (1993) and Charlton, Salpeter \& Linder (1994).
Each galaxy has an exponential inner disc and an outer extension where
the column density declines as a power law with radius, assumed as galaxy
discs are generally found to be exponential while absorbers roughly obey a power-law
column density distribution at lower column densities (see Appendix~\ref{appendixa}).
In reality a smoother
transition probably occurs, and some evidence for such a transition has been
seen by Hoffman et al. (1993).
The radius at which each
HI disc changes from exponential to power law decline was defined previously
(Linder 1998) in terms of the disc ionization edge, as little is known about
this switching radius.  Since the ionizing background varies here, however, it
makes more sense to define this switching radius in terms of the galaxy disc
scale length.  Assuming the switch from exponential to power law occurs in
each galaxy at a radius of
four HI disc scale lengths, similar results are seen in the absorber counts
arising as compared to absorber counts simulated using the previously defined switching
radius.

The galaxies
are chosen to have visible properties based upon observed distributions of
galaxy parameters, such as a Schechter luminosity function and a flat surface
brightness distribution (McGaugh 1996).	 The HI disc scale lengths ($h_{HI}$)
are assumed to
be proportional to the $B$ scale lengths ($h_B$), where $h_{HI}=1.7h_B$ to start,
as this was previously found to give rise to reasonable absorber counts in 
Linder (1998).
The gaseous properties of the
galaxies can thus be related to the visible properties.  This makes sense as galaxy
discs are generally somewhat larger in HI than in optical images, although
the HI sizes of galaxy discs also vary depending upon the location within a
cluster (Cayatte et al.~1994) so that the average ratio of $h_B/h_{HI}$ is
uncertain.
Each galaxy is placed in
a cube of space, where the positions are chosen to be clustered as described
in Linder (2000) using a fractal type
method based upon Soneira
\& Peebles (1978).  Random lines of sight through the box can then be simulated,
and an absorption line is assumed to arise when these lines of sight intersect
any disc or outer extension.  Neutral column densities for absorbers are
found by integrating
the HI density along the line of sight.

Each galaxy disc has an ionization edge, or radius beyond which no layer of 
neutral gas remains.  The vertical ionization structure of the gas is modelled
as in Linder (1998) which is similar to the model in Maloney (1993).  Inside of
this ionization radius, the gas is assumed to have a sandwich structure, where
the inner shielded layer remains neutral and has a height ($z_i$) determined by
equation (6) in Linder (1998).  The gas above height ($z_i$) and beyond the 
ionization radius is assumed to be in ionization equilibrium where $\alpha_{rec}
n_{tot}^2 = \zeta n_H$, for a highly ionized hydrogen gas with total (neutral 
plus ionized) density $n_{tot}$
and neutral density $n_H$.  The recombination coefficient is
$\alpha_{rec}=2.42\times 10^{-13}$ cm$^3$ s$^{-1}(T/10,000
\mbox{K})^{-0.75}$ and the gas temperature $T$ is assumed to be 20,000 K.
The  frequency- and direction-averaged ionization rate $\zeta$ is determined at 
each point in space while converging
numerically upon the ionization edge (the minimum disc radius where ($z_i=0$))
in a self-consistent manner, and $\zeta$ is calculated as described below.

The intensity of ionizing radiation is allowed to vary at each point in space
within the box.	 
The ionization rate $\zeta$, which was previously assumed to be constant in
Linder (1998; 2000),
is recalculated here at each point in space, where $\zeta = \zeta_b +
\zeta_{gal}$ and $\zeta_b$ is the contribution from quasars.  The value of
$\zeta_{gal}$ can be recalculated at any point within the box based upon the
flux from the surrounding galaxies.  The value of $\zeta_b=3.035 \times 10^{-14}$
s$^{-1}$ is assumed from the
calculation of Dav\'e et al.~(1999) at $z=0$ based upon spectra from Haardt \&
Madau (1996).

Bland-Hawthorn \& Maloney (1999) modelled the escape of ionizing photons from our galaxy
by extrapolating from a calculation of the ionizing photon surface density at the
Solar Circle using nearby O stars by Vacca, Garmany, \& Shull (1996).  
Bland-Hawthorn (1998) gives a simple estimate of the number of ionizing photons
escaping from the Galaxy, where $\varphi = 2\times 10^{10}e^{-\tau}r_{kpc}^{-2}cos^{0.6\tau+
0.5}\Theta$ phot cm$^{-2}$ s$^{-1}$ as in the first equation of their Appendix,
at some point with distance $r$, where $\Theta$ is the angle from the galactic pole,
and $\tau$ is the Lyman limit optical depth.
In order to add such a radiation field to $\zeta$ in the units above, we need a
frequency-averaged quantity which also considers the ionization cross section for
hydrogen, where $\sigma_H=6.3\times 10^{-18}(\nu/\nu_0)^{-3}$
at frequency $\nu$, and $\nu_0$ is the Lyman limit frequency.  Thus, for example,
the contribution from our galaxy to $\zeta$ at some point at a distance $r_{kpc}$
from its center would be 
\begin{equation}
\zeta_{gal,MW} = \int_{\nu_0}^\infty \varphi_{\nu} \sigma_H d\nu=\varphi\langle\sigma_H\rangle,
\end{equation}
where $\varphi=\int_{\nu_0}^\infty \varphi_{\nu}d\nu$.  The mean value for $\sigma_H$
is weighted by assuming that $\varphi_{\nu}\propto \nu^{-\alpha_s},$ so that
\begin{equation}
\zeta_{gal,MW}=\varphi \sigma_H(\nu_0) \left(\frac{\alpha_s-1}{\alpha_s+2}\right).
\end{equation}
We assume $\alpha_s=2.5$, which gives $\langle\sigma_H\rangle = \sigma_H(\nu_0)/3$, while
Sutherland \& Shull (1999) prefer a similar $\alpha_s\sim 1.9$ to $2.2$ for starburst
galaxies, which would
result in $\langle\sigma_H\rangle \sim \sigma_H(\nu_0)/4$.

Bland-Hawthorn \&
Maloney (1999) assumed that our galaxy has an axisymmetric exponential disc where
the ionizing photon surface density $n_d(r)=n_0e^{-r/h}$ for disc scale length $h$
and central ionizing photon surface density $n_0$.  Integrating $n_d(r)$ over the disc area
out to an infinite radius gives a number of ionizing photons which is proportional
to $h^2$, while the ionizing photon surface density can also be expressed as an ionizing
surface brightness ($\mu$) where $n/n_0=100^{(\mu_0-\mu)/5}$.
Thus we extrapolate the formula above
to other galaxies by correcting for variations in
central surface brightness ($\mu_0$) and disc scale length for galaxy $i$
as compared to Galactic (MW) values, by assuming that the ionizing scale lengths and
central surface brightnesses are proportional to the B values.
The galaxy contribution to $\zeta$ for some galaxy $i$ becomes
\begin{equation}
\zeta_{gal,i}=
\varphi\left(\frac{\sigma_H}{3}\right)
10^{(\mu_{0,MW}-\mu_{0,i})/5}\left(\frac{h_{B,i}}
{h_{B,MW}}\right)^2.
\end{equation}
At any point in space at which we wish to calculate $\zeta$ we sum the contributions
from all the galaxies in the simulation, so that
\begin{equation}
\zeta_{\rm{gal}} = \sum_i \zeta_{gal,i},
\end{equation}
where galaxy $i$ is at a distance $r$, and $\Theta_i$ 
is calculated for each galaxy as in Appendix~\ref{appendixb}.

When calculating the ionizing intensity seen by the outer part of a galaxy,
extinction from gas and dust is
important in shielding absorbing gas from being ionized by the inner parts of
the galaxy. Although some neutral gas must remain around galaxies which is often seen
to give rise to absorption, the galaxy itself still may be the most important contributor
to ionizing the gas in its outer parts.
To start we assume that each galaxy disc is flat
within two HI scale lengths
and then warped by ten degrees beyond that value (Briggs 1990),
so that the outer parts of the disc
are exposed to some ionizing radiation which escapes from the inner regions.

A value of $\tau=2.8$ is preferred by Bland-Hawthorn (1998) when modelling our Galaxy,
although the preferred value could be different as a result of an error (Bland-Hawthorn
\& Maloney 2001).
On the other hand
it has been difficult to detect any dust extinction in low surface brightness (LSB)
galaxies (for example O'Neil, Bothun \& Impey 1997).  Little is known about the
gas-to-dust ratio in LSB galaxies, although the total extinction is likely to be higher in
high surface brightness (HSB) galaxies.
Thus it might make sense to assume that $\tau_i$ is related to central surface
brightness for the simulated galaxies.  Thus a linear relationship was assumed where
$\tau=2.8$ for a galaxy with a Freeman surface brightness value of $\mu_{0,i}=21.65$,
and $\tau=0$ for $\mu_{0,i}=25$, so that
$\tau_i=-0.836(\mu_{0,i}-25)$.

Shadowing, or extinction from gas between an absorber and a given ionizing source,
is not taken into account as this would require substantially more computing
time.  This is a reasonable approximation in the sense that the universe is 
optically thin at redshifts $\ltorder 2$ (Haardt \& Madau 1996) Shadowing could
make the calculated $\zeta$ values slightly lower in some cases, although typically the 
second closest galaxy to an absorber contributes only a few percent of the ultraviolet
flux that the closest galaxy contributes.  

Starburst galaxies are not treated as emitting differently from other galaxies within
the simulation, although the galaxy population is chosen within the simulation to be
consistent with the observed optical galaxy luminosity function
and thus does not exclude the existence
of such objects.  Some of the more luminous galaxies may have a clumpy distribution of
dust, which might allow for a larger fraction of their ionizing photons to escape as
compared to other galaxies.  There could be an additional population of infrared selected
galaxies that would not be included within the luminosity function simulated here.
Although these galaxies also might make some contribution to the ionizing background,
a fairly small fraction of ultraviolet photons are thought to escape from them
(Goldader et al. 2002).

Assuming the conversion between the intensity at the Lyman limit and the one-sided flux seen
by galaxies, as defined in Tumlinson et al.~(1999), the first simulation, as illustrated in
the figures, gives
rise to a number of Lyman limit absorbers per unit redshift $(dN/dz)_{0,LL}=3.67$ when the number
density of galaxies is adjusted to produce $(dN/dz)_0=24.3$ for forest absorbers as in Bahcall
et al.~(1996).  The number density of galaxies used here is found to remain consistent with
observed galaxy luminosity functions as discussed in Linder (1998).
The observed values for  $(dN/dz)_{0,LL}$
tend to be $\sim 1$ or lower (Lanzetta, Wolfe, \& Turnshek 1995b; Storrie Lombardi et al.~1994;
Stengler-Larrea et al.~1995).  Adjusting the switching radius does not substantially change
the number of Lyman limit absorbers, as these absorbers still arise largely in the exponential
parts of the discs.  However, if the scale length ratio $h_{H1}/h_{B}$ is decreased to $1.2$
then a more realistic number of Lyman limit absorbers arises as seen in Table~\ref{table2}.  
In this case an extra
population of weaker absorbers would be needed in order to produce the observed $(dN/dz)_0$.
However most Ly$\alpha$ absorbers, including even the weakest ones, are found to trace the
large scale galaxy distribution, so that they are likely to arise in gas which is about as
close to galaxies as that in the first simulation.  However this absorbing gas could be
higher above the planes of galaxy discs and thus be even more highly ionized, as in the fifth
simulation shown in Table~\ref{table2}.  Alternatively the additional absorbers could behave more like
randomly distributed points, in which case they would have a distribution of ionization rates
similar to that for the first simulation, as illustrated in Fig.~\ref{zetadist}.

Another possibility is that Lyman limit absorbers arise only around luminous, high surface
brightness galaxies.
Observers have also questioned whether galaxies which are low in luminosity and/or surface
brightness give rise to Lyman limit absorption (McLin, Giroux, \& Stocke 1998; Steidel 1995;
Steidel et al.~1997; Bergeron \& Boiss\'e 1991),
although such objects are being found to give rise to stronger damped Ly$\alpha$ absorption
(Cohen 2001; Turnshek et al.~2000; Bowen, Tripp, \& Jenkins 2001).
Some possible explanations include gas being blown out more easily from less massive
galaxies as suggested by McLin et al.~(1998), differences in the column density profiles in the
outer parts of LSB galaxies compared to what is simulated here, a slightly
steeper neutral column density distribution compared to the $N_{HI}^{-1.5}$
assumed here, and/or a galaxy surface brightness
distribution which allows for fewer moderate to large sized LSB galaxies compared to the flat
surface brightness distribution simulated here.
Other possible scenarios and implications for the nature of
Ly$\alpha$ absorbers will be discussed further
in a future paper.

\section{Ionization Rate Fluctuations}

\begin{figure}
 \vspace{0.7cm}
\includegraphics[width=84mm]{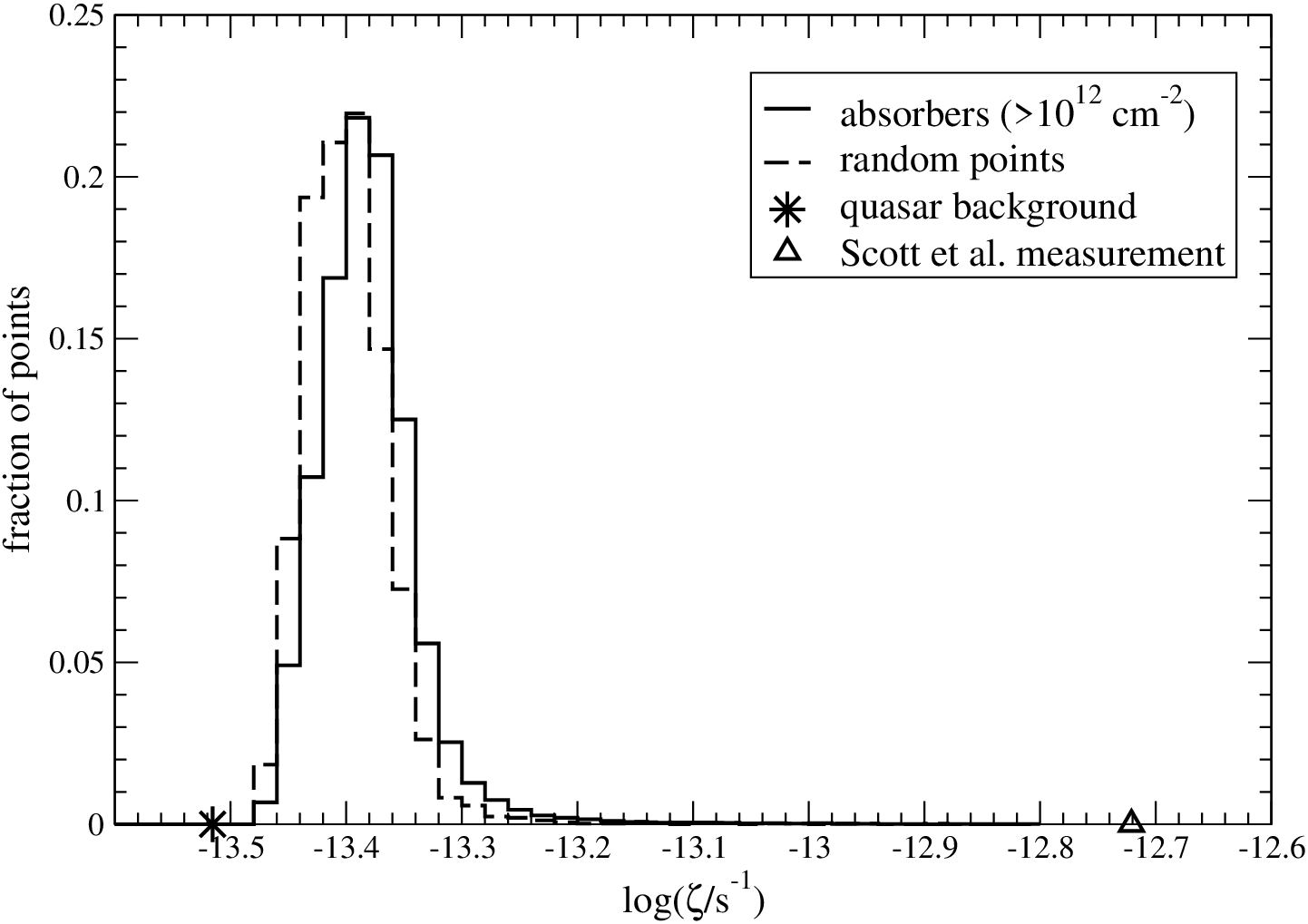}
\label{zetadist}
 \caption{A distribution of values for the direction and frequency averaged
ionization rate $\zeta$ is shown for simulated Ly$\alpha$ absorbers with
$N_{HI}>10^{12}$ cm$^{-2}$ (solid line) and for randomly chosen points within
the simulation (dashed line).  Also shown (as *) is the assumed minimum background
value due to quasars based upon Dav\'e et al. (1999) and Haardt \& Madau (1996).
The triangle indicates the value for $z<1$ found by Scott et al. (2002) using proximity
effect measurements.
}
 \label{zetadist}
\end{figure}

A further understanding of the fluctuations in the ionizing background
intensity due to galaxies is important for numerous reasons.  First
of all, the fluctuations need to be understood as a source of uncertainty
in measuring the overall background intensity.
Histograms of values for the ionization rate $\zeta$ are shown in Fig.~\ref{zetadist}
for the first simulation, 
and are shown
both for Ly$\alpha$ absorbers ($N_{HI}>10^{12}$ cm$^{-2}$) and for random points in the box.
The histograms are shown along with the $z=0$ quasar value from Dav\'e et al.~(1999), based
upon Haardt \& Madau (1996) spectra, which is the assumed minimal value for
$\zeta$.

Absorbers see higher $\zeta$ values somewhat more often than randomly chosen
points, as absorbers arise on average closer to galaxies.
Overall the ionization rate seen
by Ly$\alpha$ absorbers at $z=0$ varies over about a factor of about two in this simulation.
None of the absorbers or random points
are exposed to ionization rates which are within about ten percent
of the minimal assumed background, although extinction from gas between the
absorbers and their ionization sources is not
modelled here, so that in reality there may be a few.
Most often both the random points and absorbers have $\zeta$ values which are
about a factor of 1.4 larger than $\zeta_b$.  Varying the spectral index of the contribution
from these nearby galaxies to $\alpha_s\sim 2$ would simply move this peak to $\zeta\sim 1.3\zeta_b$.
Points with the largest values of
$\zeta$ tend to arise at quite low impact
parameters from the centres of galaxies, as will be seen in Fig.~\ref{zetavsenv}.

Also shown in Fig.~\ref{zetadist} is the
measurement from Scott et al.~(2002) for $z<1$.  This measurement has error
bars which are large compared to the range of $\zeta$ values shown in this
plot, and the lower error bar would be at a slightly higher value than the peak
of the plotted histogram, although an extra contribution from star-forming
galaxies would move the calculated peak within the measurement error bars.
While the galaxies are simulated
here at $z\sim 0$, the ionization rate is seen to evolve only by $<0.2$ orders
of magnitude between the Scott et al.~(2002) measurements at $z<1$ and $z>1$.
According to
the redshift evolution model they fit, where the background intensity evolves
as $(1+z)^{0.017}$, there would be even less difference expected between their
measurement for $z<1$ and what would be expected at $z\sim 0$.
This could mean that more ionizing radiation escapes from galaxies
than is assumed here, although it is more likely that this power law model does
not describe the evolution in the ionizing background very well down to
$z\sim 0$.  However, an additional uniform contribution could
also come from galaxies at redshifts $\ltorder 2$, as the universe is optically
thin to ionizing photons at these redshifts.  Such a contribution would likely 
be $\sim\zeta_b$, as calculated for star-forming galaxies 
(Giallongo et al.~1997; Shull et al.~1999; Bianchi et al.~2001).  Another possible
contribution could come from gas in the intragroup medium (Maloney \& Bland-Hawthorn
1999; 2001).
On the other hand, the other previously mentioned
low redshift measurements also
tend to be consistent with $\zeta$ values closer to the assumed $\zeta_b$, so the value
at $z=0$ need not be as large as the Scott et al. (2002) measurement.

While the background intensity
detected using the proximity effect is likely to be the most common value
in a peaked distribution like the one found here, other values could be
seen, for example, when making measurements of the ionization rate around a galaxy in an
unusual environment. 

\begin{figure}
 \vspace{0.7cm}
\includegraphics[width=84mm]{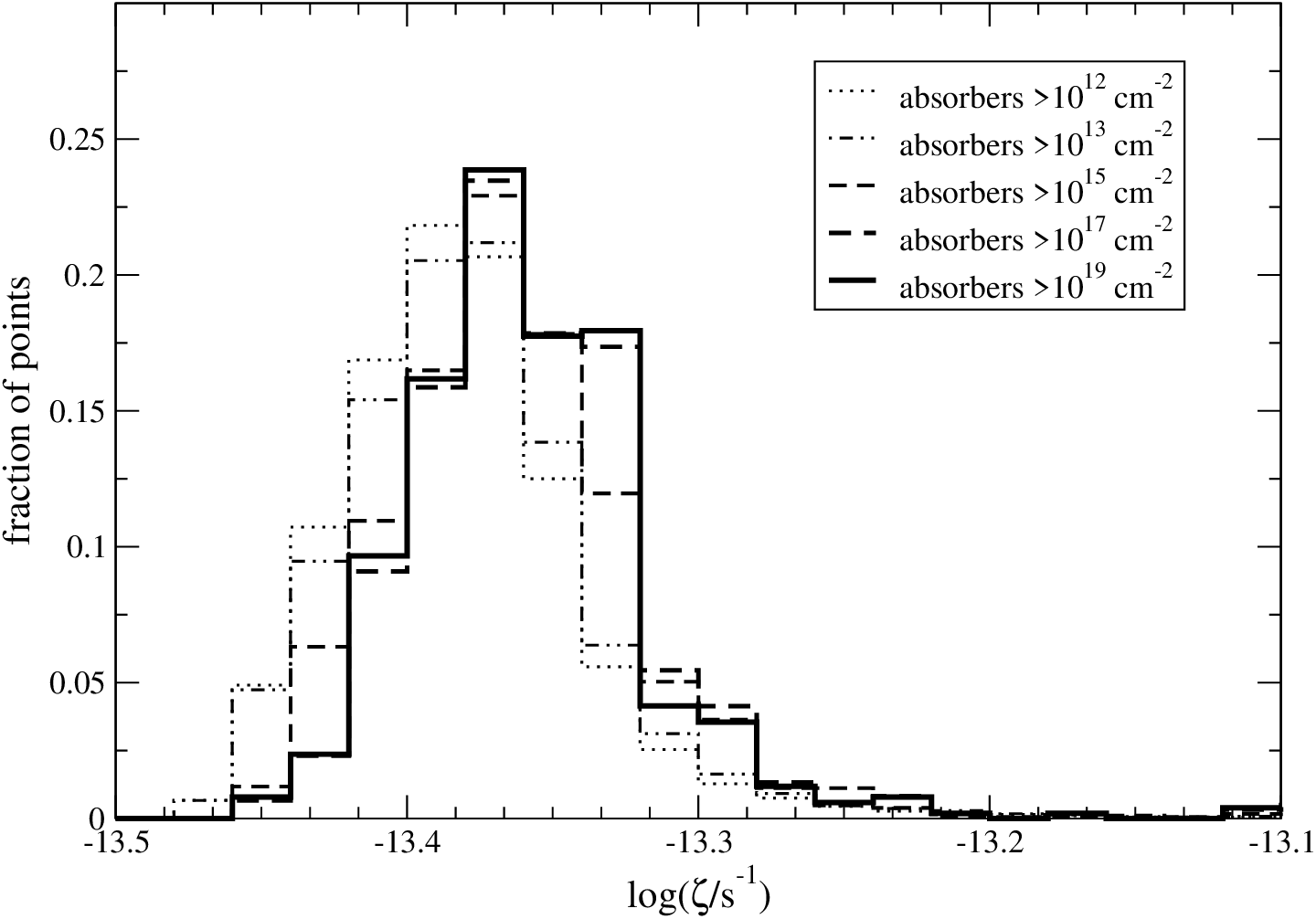}
 \caption{A distribution of values for the
ionization rate $\zeta$ is shown for simulated Ly$\alpha$ absorbers with
varying limiting neutral column densities.  Values of $\zeta$ are binned for
absorbers having
$N_{HI}>10^{19}$ cm $^{-2}$ (solid line),
$>10^{17}$ cm $^{-2}$ (long-dashed line),
$>10^{15}$ cm $^{-2}$ (dashed line),
$>10^{13}$ cm $^{-2}$ (dot-dashed line), and
$>10^{12}$ cm $^{-2}$ (dotted line).  Stronger absorbers arise on average
closer to luminous galaxies, so that their $\zeta$ values are typically higher.
}
 \label{zetadistlim}
\end{figure}

\begin{table}
\caption{Ionization Rates for Varied Limiting Absorber Column Densities}\label{table1}
 \begin{tabular}{@{}ccc}
  \hline
  $N_{HI,min}/cm^{-2}$
        &$\overline{\log (\zeta/s^{-1})}$ & $\sigma$\\
  \hline
 $10^{12}$&-13.38&0.06\\
 $10^{13}$&-13.38&0.10\\
 $10^{15}$&-13.36&0.33\\
 $10^{17}$&-13.36&0.58\\
 $10^{19}$&-13.36&0.63\\
  \hline
 \end{tabular}

\medskip
For Ly$\alpha$ absorbers with varying limiting neutral column densities, $N_{HI,min}$,
the mean values for $\log\zeta$ and standard deviations are shown for the first simulation
as also illustrated in Fig.~\ref{zetadistlim}.  Absorbers with higher $N_{HI}$ appear
to have more higher $\zeta$ values in the Figure.  Here it can be seen that the mean
$\log \zeta$ values change only slightly with limiting $N_{HI}$, but the $\sigma$ values
become much larger, indicating more substantial tails of high $\zeta$ values.
\end{table}

Distributions of $\zeta$ values for absorbers are shown 
in Fig.~\ref{zetadistlim}, where the limiting neutral column density for absorbers is
varied.  It can be seen that more high $\zeta$ values arise for stronger absorbers,
as these arise on average closer to galaxies.  The average $\log \zeta$ value increases
only slightly with limiting $N_{HI}$ however, but the distributions become much less
strongly peaked due to larger tails of high $\zeta$ values, as shown in Table~\ref{table1}.

\begin{figure}
 \vspace{0.7cm}
\includegraphics[width=84mm]{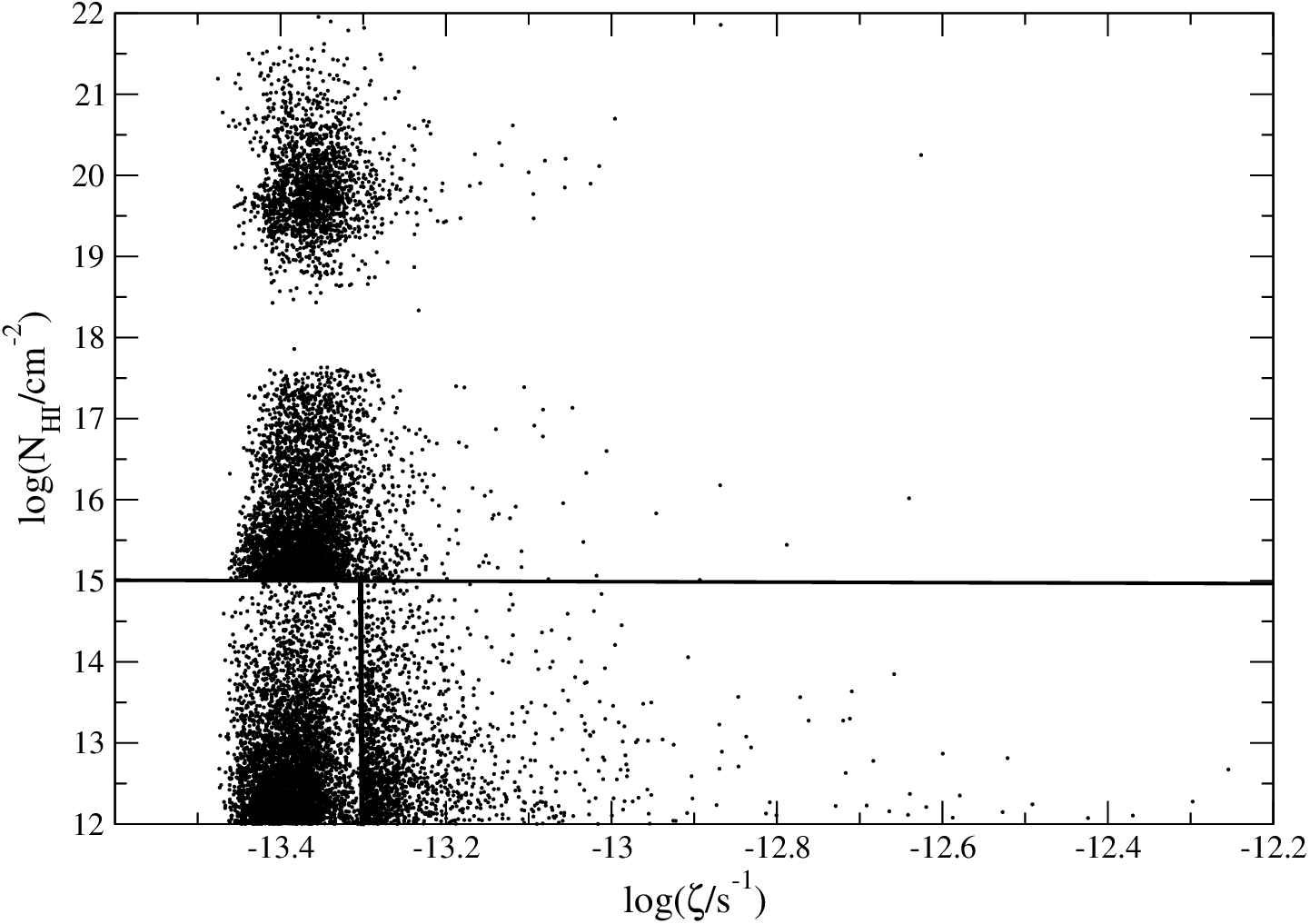}
 \caption{Values of the neutral column density are plotted versus ionization
rate for simulated Ly$\alpha$ absorbers. Absorbers are shown as arising from 15,000
lines of sight in the region where
$N_{HI}<10^{15}$ cm$^{-2}$ and  $\zeta > 10^{-13.3}$ s$^{-1}$, and
ten percent of these simulated points
are shown in the region where $N_{HI}<10^{15}$ cm$^{-2}$ and  $\zeta < 10^{-13.3}$
s$^{-1}$ which would otherwise be further saturated.  Absorbers are shown as arising
from 60,000 lines of sight through the box where $N_{HI}>10^{15}$ cm$^{-2}$.
Fairly high values of $\zeta$ can
be seen for any value of $N_{HI}$.  Few points arise with $N_{HI}\sim 10^{18}$
cm$^{-2}$ because absorbers change from neutral to highly ionized at a slightly
larger $N_{HI}$ value, as seen in galaxy disc ionization edges.}
 \label{zetavsnh}
\end{figure}

Values of $\zeta$ are plotted versus neutral column density ($N_{HI}$) for absorbers in
Fig.~\ref{zetavsnh}.  Different numbers of simulated points are plotted in different
regions of the plot in order to reduce saturation for low $\zeta$ values and to show
more detail for absorbers with high $N_{HI}$.  Again the minimum assumed $\zeta=\zeta_b$
produces a cutoff on the left side of the
plot.  Few points are seen with $N_{HI}\sim 10^{18}$ to $10^{19}$, which is related to the
galaxy disc ionization edges where $N_{HI}$ falls off quickly with increasing radius around
these values.  It can be seen
that absorbers with a wide range of $N_{HI}$ can be exposed to a
wide range of ionization rates.  Even more absorbers with a wide range of higher
$\zeta$ values would be seen for a wide range of $N_{HI}$
if more ionizing photons were allowed to escape from
galaxies in the simulation.  This happens even though one might expect absorbers exposed
to large $\zeta$ to be ionized away or seen only at lower values of $N_{HI}$.  High $N_{HI}$
values with high $\zeta$ still arise, as the galaxies are simulated with a full range of disc
inclinations.

Occasionally voids are reported in the Ly$\alpha$ forest (Crotts 1987; Dobrzycki \&
Bechtold 1991; Cristiani et al.~1995), and there is the
possibility that variations in the ionizing background might contribute to such
voids.  However at this time numbers of voids beyond what would be expected for
a random absorber population have not yet been detected for voids smaller than
the box size simulated here (Williger 2002).

\section{Model Parameter Variations}

\begin{table}
\caption{A Summary of the Simulations}\label{table2}
 \begin{tabular}{@{}lcccc}
  \hline
 Description&$(dN/dz)_0$&$(dN/dz)_{0,LL}$
       &$\overline{\log (\zeta /s^{-1}) }$ & $\sigma$\\
  \hline
  1: As in text &25.93&3.91&-13.38
        &0.06\\
  2: $\tau=2.8$&27.43&4.01&-13.43
        &0.05\\
  3: $\tau=3.5$&28.85&4.09&-13.47
        &0.03\\
  4: $\tau=1.5$&21.72&3.69&-13.23
	&0.10\\
  5: warp$=25^o$&24.47&3.69&-13.36
        &0.07\\
  6: $h_{21}/h_B=1.2$&14.06&0.90&-13.44
        & 0.04\\
  \hline
 \end{tabular}

\medskip
The first simulation is as described in the text and illustrated
in the Figures.  Shown for each simulation in the table are the
parameters which were varied (where each simulation is otherwise
the same as the first simulation), and the numbers of absorbers $>10^{14.3}$
cm$^{-2}$ and Lyman limit
absorbers arising per unit redshift in a simulation with 12590 galaxies
in a 28.9 Mpc cube, and the mean values
for the logarithm of the
ionization rate $\zeta$ for Ly$\alpha$ absorbers, and the
standard deviation for the distribution of $\log\zeta$.
\end{table}

Several model parameters were varied in further simulations
as described in Table~\ref{table2}.  The
number of Lyman alpha absorbers having $N_{HI}>10^{14.3}$ cm$^{-2}$, the
number of Lyman limit systems, and the mean and standard deviation $\sigma$
for $\log \zeta$ are shown for each simulation.  In each case the mode
for the distribution of $\log \zeta$ is very close in value to the mean.
In cases where $\sigma$ is relatively large, there tends to be a more
substantial
tail of points having high $\zeta$ values.
The optical depth $\tau$ was varied
in order to explore the uncertainty range in the fraction of
ionizing photons escaping from galaxies.  The value of $\tau=2.8$,
preferred for our Galaxy by Bland-Hawthorn (1998), corresponds to a
direction-averaged escape fraction of $\sim 2$ percent of the ionizing
photons (Bland-Hawthorn \&
Maloney 2001).  In the first simulation, it is assumed that $\tau$ is
dependent upon galaxy central surface brightness as discussed above,
while in the second simulation we assume $\tau=2.8$ for all galaxies.
In the third simulation  $\tau=3.5$ is used for all galaxies
in order to produce an escape
fraction of $\sim 1$ percent, while $\tau=1.5$ is used for all galaxies
in the fourth simulation, giving an escape fraction $\sim 10$ percent.
A similar distribution for $\zeta$ arises when the value $\tau=2.8$ is
assumed for all galaxies as compared to the first simulation.
The mean and mode values of  $\log \zeta$ correspond to $1.38\zeta_b$ in the
first simulation, $1.23\zeta_b$ in the second simulation, $1.11\zeta_b$
where  $\tau=3.5$, and $1.95\zeta_b$ where $\tau=1.5$.  Thus the
uncertainty in the fraction of ionizing photons escaping from galaxies
means that normal galaxies could contribute between ten percent of the
quasar background and an amount about equal to the quasar background.

The typical disc warping angle, which is also rather
uncertain, was also varied as shown in the fifth simulation in
Table~\ref{table2}.  Bland-Hawthorn (1998) has suggested, for example, that there may
be a selection bias against detecting highly warped discs in HI because
they become more highly ionized by the stars within the galaxy.
When the disc warping angle is increased to $25^o$ then a slightly
higher mean $\log \zeta$ is seen, as more ionizing photons escape
higher above the plane of a disc.  However the absorbers even in this
case are not generally far above the plane of the disc, as the disc is only
warped beyond two HI scale lengths.

The distributions in ionization rates seen in the simulations here are generally
quite strongly peaked, although there may be more absorbers with high $\zeta$ if
there is substantial absorbing gas above the planes of galaxy discs.
The contribution
to the ionization rate varies by a factor of $45$ for the Bland-Hawthorn (1998) model
between polar angles of $0^o$ and $80^o$ at some distance from a galaxy with $\tau=2.8$.
An additional simulation was done where
absorbers arise in galaxy haloes with column density profiles obeying equation
(24) from Chen et al.~(1998).  While the absorbing gas is not modeled in this
case, the distribution of ionization parameters is found to be similar to that
seen in the fifth simulation (where the warping angle $=25^o$).  
Even in this case, however, or for randomly chosen points,
or in any case where absorbers arise in random directions relative to galaxies, few of
the points or absorbers will arise close to the galactic poles.

The sixth simulation was done with a reduced  $h_{21}/h_B=1.2$ in order to illustrate
a scenario with a more realistic number of Lyman limit absorbers.  In this case the
mean $\log \zeta$ is slightly lower compared to the first simulation because the simulated
absorbers arise typically somewhat closer to the centres of galaxies within the warped
outer discs, so that they arise closer to the planes of the discs.  In order to make
such a scenario realistic, however, an additional population of low column density
absorbers would be needed to account for the observed $(dN/dz)_0$ as in Bahcall et
al. (1996).  The additional absorbers could either arise far from luminous galaxies
and thus behave more like the random points illustrated in Fig.~\ref{zetadist}, or
they could arise higher above galactic planes where they could be
more highly ionized.

A simulation was also run at redshift one by decreasing the box size by a
factor of $(1+z)$ and assuming $\zeta_b=3.33\times 10^{-13}$ s$^{-1}$, again based upon
the calculations of Dav\'e et al.~(1999) and Haardt \& Madau (1996) at $z=1$.
Any inconsistency
might indicate some galaxy luminosity evolution and/or require more photons to escape
from galaxies at $z=1$, although the $\zeta$ distribution was found to be peaked at
$3.7\times 10^{-13}$ s$^{-1}$, which is between the Scott et al.~(2002) measurements
for $z<1$ and $z>1$.

\section{Relationship to Galaxies}

Understanding how the ionizing background varies with galactic environment will
be important for understanding what kinds of objects give rise to
Ly$\alpha$ absorption, and in which environments Ly$\alpha$ absorption is
most likely to arise.
Tripp et al.~(1998) suggest that no absorption is
found near a galaxy cluster due to increased ionization.
Fluctuations in the ionizing background may also allow
for substantial variations in the environments affecting galaxy formation
processes.  For example, it has been
suggested that dwarf galaxies may form less easily in an
intense radiation field (Tully et al.~2002; Efstathiou 1992; Quinn, Katz, \&
Efstathiou 1996; but see the discussion in Sabatini et al.~2003).

\begin{figure}
 \vspace{0.7cm}
\includegraphics[width=84mm]{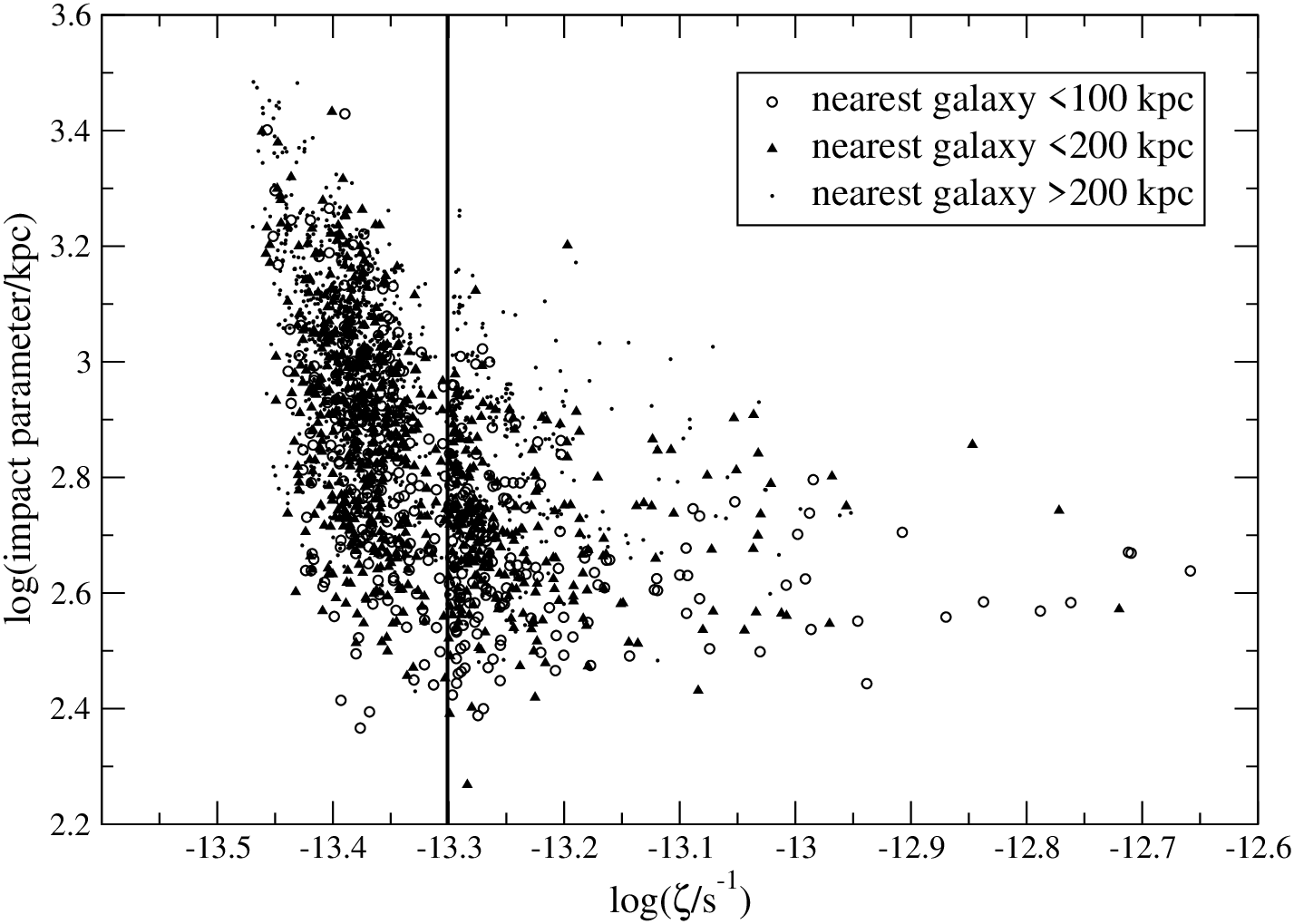}
 \caption{The average impact parameter to the nearest four galaxies having
$M_B<-16$ is plotted versus ionization rate for simulated Ly$\alpha$
absorbers with $N_{HI}>10^{12}$ cm$^{-2}$.  Circles indicate points where the
nearest galaxy having $M_B<-16$ is within 100 kpc, and triangles indicate
such a galaxy within
200 kpc.  Eight percent of the simulated points with the nearest galaxy
$>200$ kpc and fifteen percent of the other points are shown in the region
where $\zeta<10^{-13.3}$ s$^{-1}$ where the points would otherwise be further
saturated.
Absorbers with the largest $\zeta$ values tend to be in environments
which are rich in nearby galaxies.  Absorbers with lower values of $\zeta$ arise
at a wide range of averaged impact parameters.}
 \label{zetavsenv}
\end{figure}

In order to attempt to parametrize the galaxy clustering environment,
in Fig.~\ref{zetavsenv} $\zeta$ values are plotted (again for the first simulation)
versus the average distance to the
nearest four galaxies having $M_B < -16$, as observers are generally unable to
detect less luminous galaxies around even the nearest absorbers.  Again here and in the next
two figures, different numbers of simulated points are plotted for $\log \zeta<-13.3$
values in order to reduce the saturation in the plot.  The nearest four galaxies are
used because the ionizing background is likely to be somewhat higher even in a
group environment in addition to being higher in a rich cluster.  The shapes of the
plotted points give more information about the nearest single galaxy with $M_B < -16$.
The $\zeta$ values
are shown for absorbers with $N_{HI}>10^{12}$ cm$^{-2}$.
Most of the points
with $\zeta\sim\zeta_b$ are far from luminous galaxies, while those with
higher $\zeta$ arise more often closer to luminous galaxies.  However, selection
effects against LSB galaxies may make this correlation less clearly visible to
an observer.
Some higher
$\zeta$ values can arise even far from luminous galaxies, however, as it was
assumed that some ionizing radiation escapes even from dwarf galaxies which
are not assumed to be strongly clustered.  Still the presence of any luminous
galaxy may have a more important effect on the ionization rate rather than
the overall clustering environment, as large $\zeta$ values tend to arise
more often when a luminous galaxy is within 200 kpc.

\begin{figure}
 \vspace{0.7cm}
\includegraphics[width=84mm]{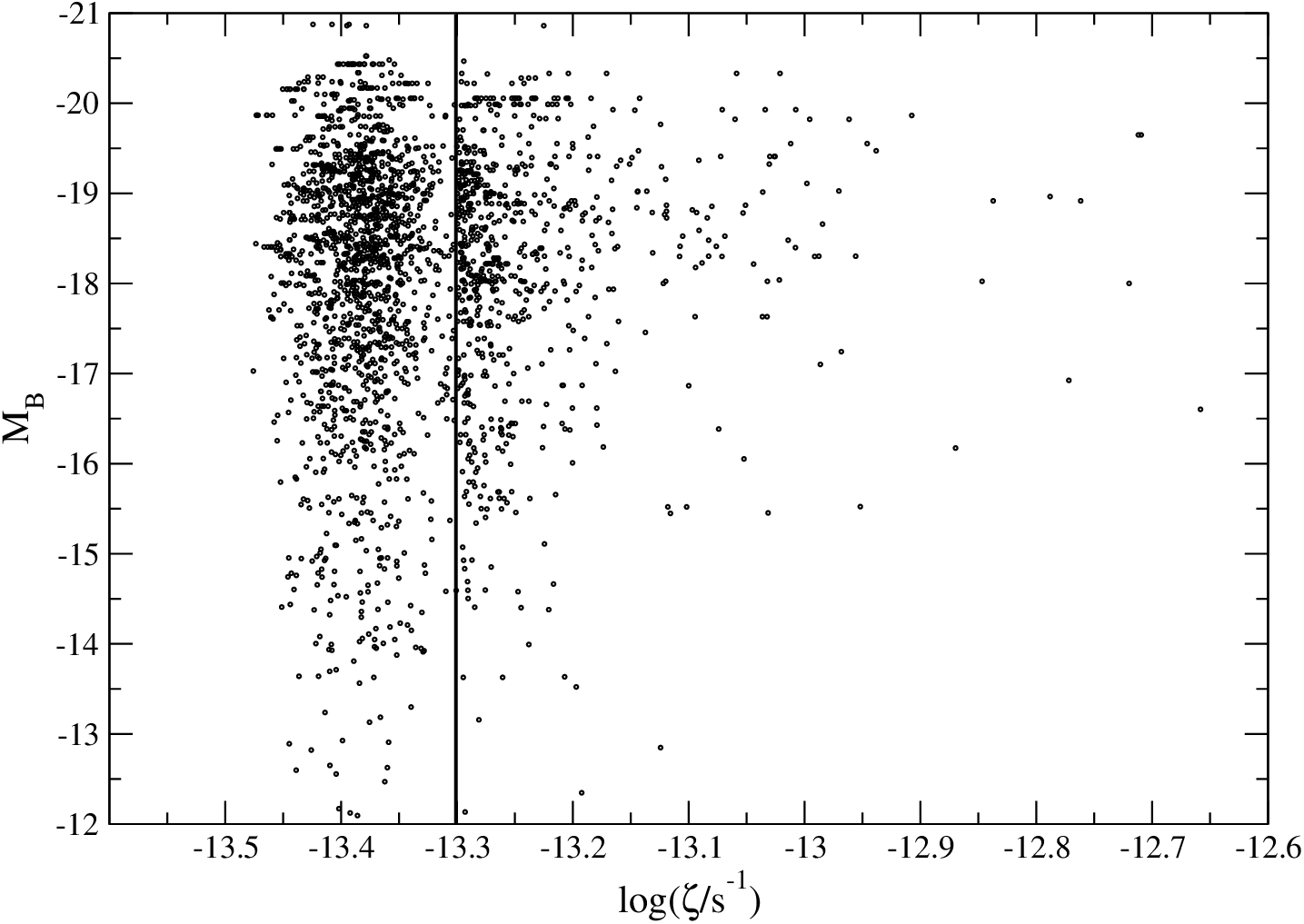}
 \caption{Absolute magnitudes are plotted for galaxies where Ly$\alpha$
absorption ($>10^{12}$ cm$^{-2}$) is known to arise in the simulation
versus ionization rate for
the absorber.  Ten percent of the simulated points are shown for $\log\zeta<-13.3$
where the figure would otherwise be further saturated.
Horizontal lines are seen because numerous absorbers can arise
close to a particular simulated luminous galaxy.  More high $\zeta$ absorbers
are seen around luminous galaxies simply because these galaxies are assumed to
have more absorbing gas around them so that more absorbers with any $\zeta$
value arise.  Fairly
high values of $\zeta$ can arise around galaxies with any $M_B$ value.
 }
 \label{zetavsmb}
\end{figure}

\begin{figure}
 \vspace{0.7cm}
\includegraphics[width=84mm]{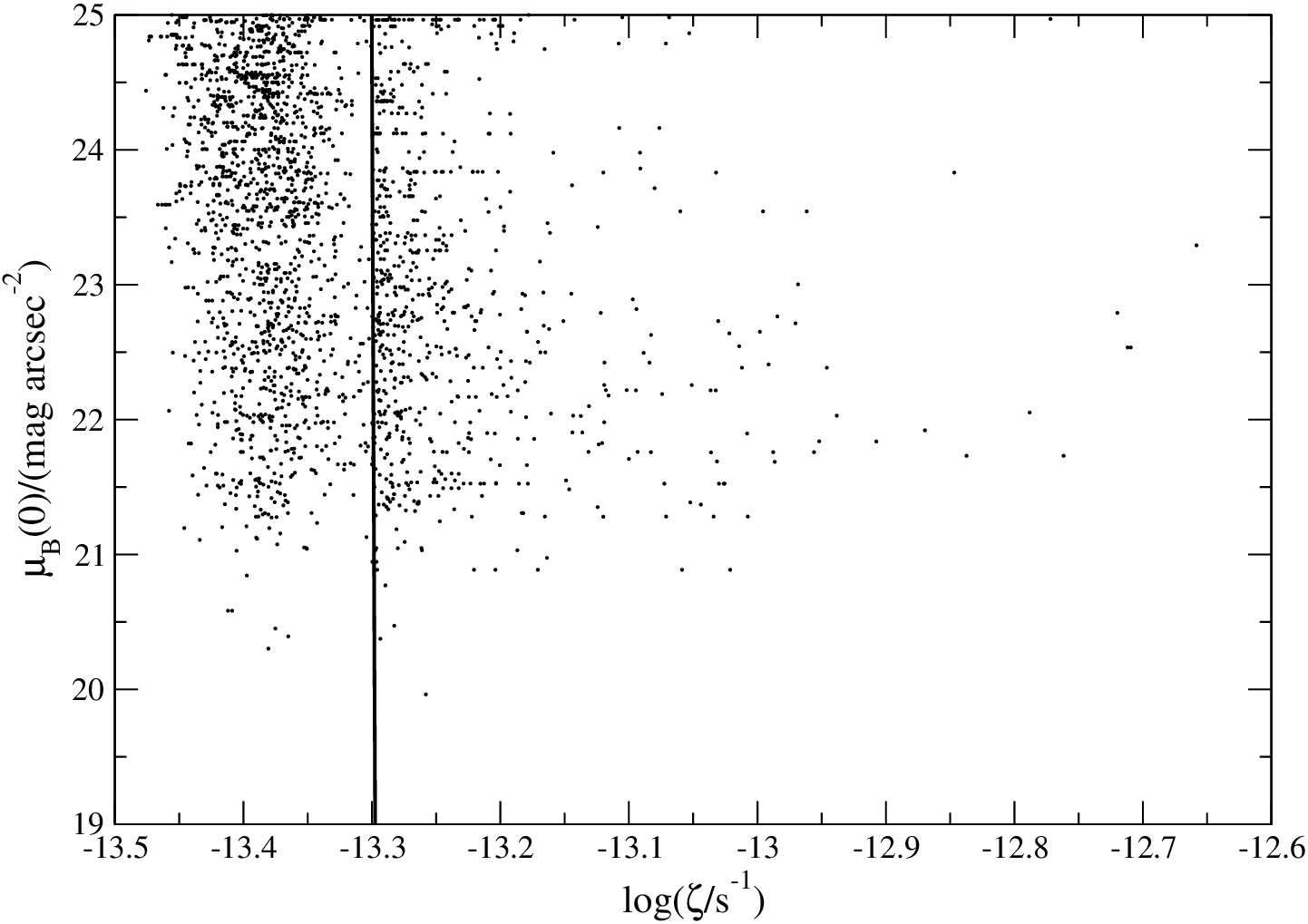}
 \caption{Central surface brightnesses are plotted for galaxies where Ly$\alpha$
absorption ($>10^{12}$ cm$^{-2}$) arises nearby versus ionization rate for
the absorber.
Ten percent of the simulated points are shown for $\log\zeta<-13.3$
where the figure would otherwise be further saturated.
High $\zeta$ values can arise around galaxies with a wide range
of central surface brightness values.
}
 \label{zetavsmuo}
\end{figure}

In Figs.~\ref{zetavsmb} and \ref{zetavsmuo},
$\zeta$ values are plotted versus galaxy luminosity and
central surface brightness for each absorber, where their associated galaxies are
known from the
simulation.  (Note that an observer might identify a different galaxy as
associated with an absorber, as discussed in Linder (2000).)
Points are seen to lie on horizontal lines in either plot,
as galaxies which are luminous or
moderately
low in surface brightness are assumed to have large absorption cross sections
for their surrounding gas and give
rise to numerous absorbers within the simulation.  It can be seen in either
plot that absorbers arising around galaxies with a wide range of properties
are exposed to a wide range of ionization rates.  Thus variations in $\zeta$
happen around particular galaxies, although these galaxies can have a wide
range of properties.  High $\zeta$ values will be seen most often in locations
where absorbers arise most often, such as those close to luminous galaxies.
Yet many more faint galaxies exist, where some absorbers with high $\zeta$ values
can also be seen, and no evidence is seen for a variation in the average
$\zeta$ values with galaxy luminosity or surface brightness.

\begin{figure}
 \vspace{0.7cm}
\includegraphics[width=84mm]{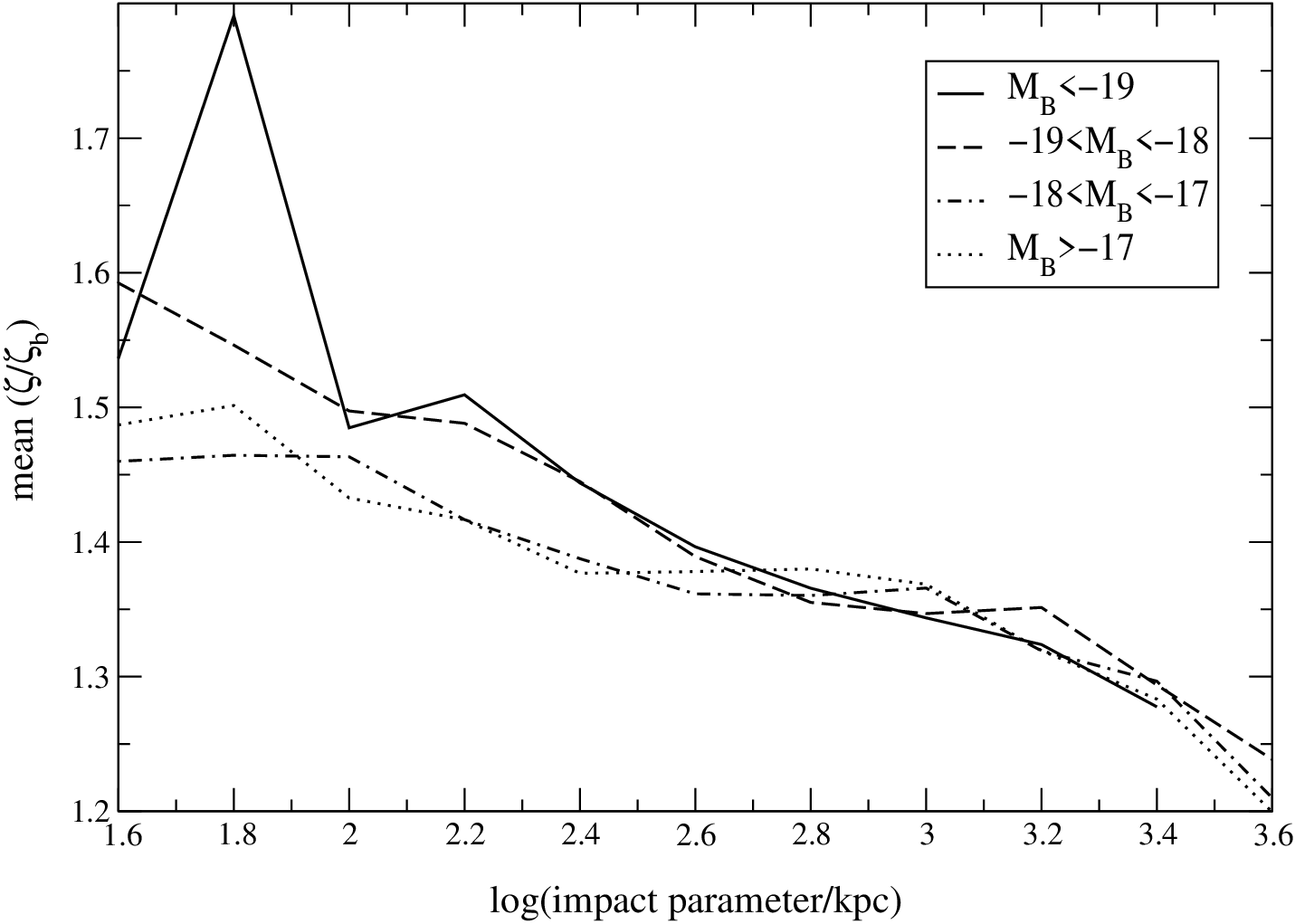}
 \caption{The average ionization rate, shown in units of the assumed quasar
background $\zeta_b$, is shown here for absorbers $>10^{12}$ cm$^{-2}$ versus
impact parameter to the nearest galaxy with $M_B>-16$.  A solid line is shown
where the nearest such galaxy has $M_B<-19$, a dashed line for $-19<M_B<-18$,
dot-dashed for $-18<M_B<-17$, and dotted for $M_B<-17$.  It can be seen that
the average $\zeta$ value for absorbers decreases with increasing impact
parameter from a luminous galaxy, and the decrease appears to be steepest
for the most luminous galaxies.  Note that the error bars based upon the standard
deviation in simulated $\zeta$ values would be smallest
for impact parameter values in the centre of the plot where $\zeta/\zeta_b\sim 1.4$,
as there are fewer data points for high and low $\zeta$ values.
}
 \label{zetabinip}
\end{figure}

The ionization rate tends to be higher close to galaxies and in regions of higher galaxy
density where Ly$\alpha$ absorbers often arise, but how much is the
intergalactic medium affected on average by ionizing radiation from galaxies?
The lowest $\zeta$ values tend to be seen when looking as far as possible
from a luminous galaxy, as can be seen in Fig.~\ref{zetabinip}.  A fall-off can be seen in the
average $\zeta$ value with impact parameter from a galaxy with $M_B<-16$, and
such a fall-off appears to be steeper for more luminous galaxies.  Again this
plot may be affected by luminous galaxies having more absorbers
with high $\zeta$ values around them simply because more absorbing gas is
assumed to be located around luminous galaxies.

\begin{figure}
 \vspace{0.7cm}
\includegraphics[width=84mm]{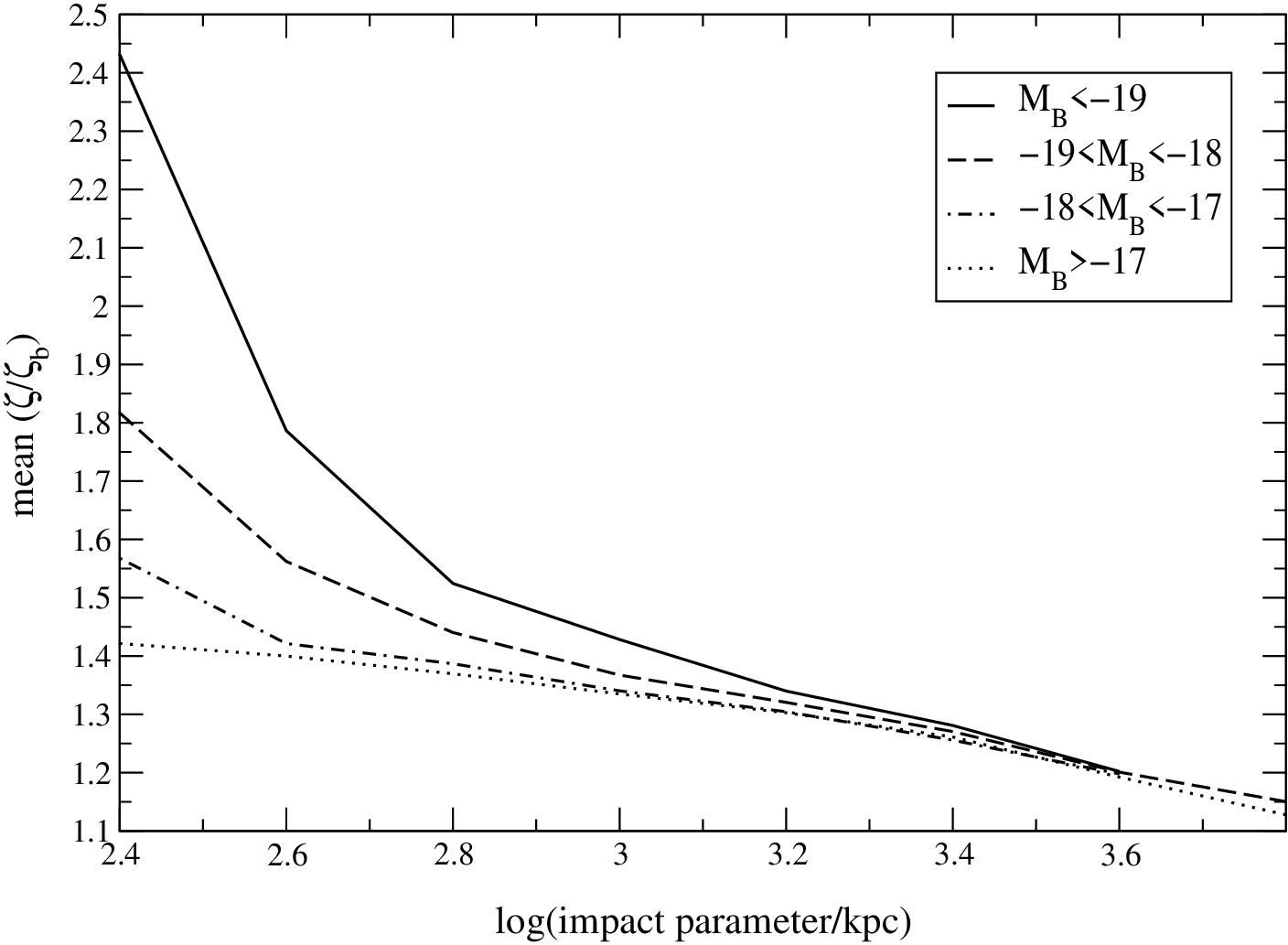}
 \caption{The average ionization rate, shown in units of the assumed quasar
background $\zeta_b$, is shown here at randomly chosen points versus
impact parameter to the nearest galaxy with $M_B>-16$.  A solid line is shown
where the nearest such galaxy has $M_B<-19$, a dashed line for $-19<M_B<-18$,
dot-dashed for $-18<M_B<-17$, and dotted for $M_B<-17$.  Compared to the
previous figure, it can be seen even more clearly here that
the average $\zeta$ value for absorbers decreases with increasing impact
parameter from a luminous galaxy, and the decrease is the steepest
for the most luminous galaxies.
}
 \label{zetabinipis}
\end{figure}

A less biased view of the ionization of the intergalactic medium can be seen
from looking at a similar plot of the average $\zeta$ value versus impact parameter
to the nearest galaxy for random points rather than absorbers, as shown in
Fig.~\ref{zetabinipis}.  While it becomes even more difficult here to simulate points
which are very close to galaxies, it can be seen even more clearly that the
intergalactic medium is more highly ionized at points which are closer to more
luminous galaxies.  While Fig.~\ref{zetabinip} is a prediction of what observers
might see if they are able to measure $\zeta$ for numerous absorbers,
Fig.~\ref{zetabinipis} is a better representation of how the ionization of the
intergalactic medium could be simulated.

\section{Conclusions}

Normal galaxies are likely to contribute at least thirty to forty percent of what quasars
do to the ionizing background of Lyman continuum photons at zero redshift,
assuming that ionizing photons escape from other galaxies in an analogous
manner to the Bland-Hawthorn (1998) model of our Galaxy where $\sim 2$ percent of
ionizing photons escape.  Allowing for some uncertainty in this direction-averaged
ionizing photon escape fraction, assuming that between $\sim 1$ and $\sim 10$ percent
of ionizing photons escape means that
the contribution to the ionizing background from normal
galaxies could be between $\sim 10$ and $\sim 100$ percent of the assumed quasar contribution.
This ultraviolet background is important for ionizing gas surrounding galaxies and within
the intergalactic medium.  This gas gives rise to Ly$\alpha$ absorption and makes
an uncertain contribution to the baryon content in the local universe due to uncertainties
in ionization intensities and mechanisms.

Distant quasars at somewhat higher redshifts will ionize the low redshift universe
in a relatively uniform manner, but ionizing radiation escaping from normal galaxies
at low redshift will result in local fluctuations in the ionizing background.
Fluctuations have been found in the ionization rate of gas around
simulated
galaxies, which will give rise to variations in the neutral gas fractions in Ly$\alpha$
absorbers with a wide range of neutral column densities.  A wide range of ionization
rates are found to arise close to galaxies having a wide range of properties, but normal galaxies
also play an important role in ionizing the more distant intergalactic medium.
Ionization rates
for absorbers are found to
be about twice as high as the quasar background on average when looking at $\sim 200$ kpc 
from a luminous galaxy, about 1.4 times the quasar background when looking at $\sim 1$ to 
2 Mpc from a luminous galaxy, and only as little as $\sim 1.1$ times the quasar background
when looking as far as possible
from luminous galaxies. 
Luminous star forming galaxies, which may contain rather clumpy dust and thus have more
complex extinction behaviour than was modelled here, will also give rise to both
local and larger scale fluctuations in the ionizing background.

Fluctuations in the ionizing background may have implications
for the formation and evolution, and our ability to detect, smaller objects located near
luminous galaxies, such as dwarf galaxies and high velocity clouds.
Such fluctuations will also give rise to variations
in galaxy disc ionization edges, which will have implications for
the column density distribution of Lyman limit
absorbers.  Fluctuations in the ionizing background may also have some implications
for the nature of Ly$\alpha$ absorbers and their relationship to galaxies.
If galaxies play an important role in ionizing the gas around them, gas which must make
some contribution to Ly$\alpha$ absorption, then further constraints may be made on models
for galaxies giving rise to absorption.  Gas which is too close to a luminous galaxy, and in
particular that located far above the plane of the disc,
will be exposed to more ionizing radiation than absorbing
gas which is ionized largely by a quasar background,
possibly reducing the number density of absorbers
that will arise in such environments.  Variations in the fractions of
ionizing radiation that escape from galaxies with various properties may be important
for determining what kinds of galaxies can give rise to absorption.
A further understanding of the distribution of ionization rates will enable us
to learn more about chemical abundances using metal absorption line systems.

The fluctuations in the ionizing background are not seen to be substantial
in the simulations here however, although most absorbers are assumed to arise fairly close to the
planes of galaxy discs.  Larger variations in the photoionization rates would
only be seen if substantial amounts of absorbing gas are concentrated above
galactic poles, although in this case the gas might include components which are
ejected from the galaxies so that collisional ionization processes might also be
important.

\section*{Acknowledgments}  We are grateful to 
R.~Dav\'e, S.~Eales, and J.~Scott for valuable 
discussions and to J.~Charlton and the referee, S.~Bianchi, for 
careful reading and helpful suggestions for improving the manuscript.

\appendix
\section{Relationship between the Column Density Profile in a Disc and the
Column Density Distribution}\label{appendixa}

Suppose absorbers arise in the outer parts of galaxy discs, where the column
density $N_{HI}$ in each disc falls off with radius $r$ as a power law
where
$N_{HI}\propto r^{-p}$.  The neutral column density distribution
 $d{\cal{N}}/dN_{HI}$
resulting from these absorbers in then $d{\cal{N}}/dN_{HI}=
 d{\cal{N}}/dr\times dr/dN_{HI} = 2\pi r \times  dr/dN_{HI}$.  Since
$dN_{HI}/dr \propto r^{-(p+1)}$ from the assumed column density profile then
 $d{\cal{N}}/dN_{HI} \propto r^{(p+2)}\propto (N_{HI}^{-1/p})^{(p+2)}$.
Thus when the neutral column density distribution in each outer disc falls
off as a power law with exponent $-p$, then the resulting neutral column density
distribution is also a power law having exponent $-\epsilon$, where
$d{\cal{N}}/dN_{HI} \propto N_{HI}^{-\epsilon} \propto  N_{HI}^{-(p+2)/p}$
and $p=2/(\epsilon-1)$.

\section{Polar Angle Calculation}\label{appendixb}

At a point $(x,y,z)$ in space where the ionization rate is calculated, each galaxy $i$
contributes emission which is seen at angle $\Theta_i\ (\leq\pi/2)$ from the galaxy's pole.
Each galaxy has a randomly simulated inclination $\theta$ between the disc plane and the
$z$-axis, thus chosen from a uniform distribution
in $\cos \theta$, and a random orientation $\phi$, where $0\leq\phi<2\pi$.  

The angle between the galaxy's rotation axis given by the vector ${\bf v}$ and
the line given by the vector ${\bf d}=(x-x_i,y-y_i,z-z_i)$ equals
\begin{eqnarray*}
\lefteqn{\min(\angle({\bf v},{\bf d}),\angle(-{\bf v},{\bf d}))} \\
 &=&
\min\left(\cos^{-1}\left(\frac{\langle {\bf v},{\bf d}\rangle}{||{\bf
v}||\cdot||{\bf d}||}\right),\cos^{-1}\left(\frac{\langle
{\bf v},-{\bf d}\rangle}{||{\bf v}||\cdot||{\bf d}||}\right)\right) \\&=&
\cos^{-1}\left(\frac{|\langle {\bf v},{\bf d}\rangle|}{||{\bf v}||\cdot||{\bf
d}||}\right) \\&=&
\cos^{-1}\left(\frac{|v_xd_x+v_yd_y+v_zd_z|}{||{\bf v}||\cdot||{\bf
d}||}\right)
.\end{eqnarray*}
 This angle is always in the interval $[0,\pi/2].$ Here
$||{\bf d}||=\sqrt{d_x^2+d_y^2+d_z^2}$.
The vector ${\bf v}$ is produced by
rotating the vector $(0,0,1)$ first by $\pi/2-\theta$ in the $xz$-plane,
which gives the vector $(\cos\theta,0,\sin\theta)$, and then rotating by angle
$\phi$ in the $xy$-plane, which gives the vector
$(\cos\theta\cos\phi,-\cos\theta\sin\phi,\sin\phi).$ Hence $||{\bf v}||=1$.

The polar angle is thus
\begin{equation}
\Theta_i = \cos^{-1} \left( \frac{|d_x\cos\theta\cos\phi-d_y\cos\theta\sin\phi+
d_z\sin\theta|}{(d_x^2+d_y^2+d_z^2)^{1/2}}\right)
\end{equation}
where
\begin{eqnarray}
d_x = x-x_i\\
d_y = y-y_i\\
d_z = z-z_i
\end{eqnarray}
for galaxy $i$ centred at $(x_i,y_i,z_i)$.

\end{document}